# Power spectral density analysis of relative comb-line phase jitter in a twin-soliton molecule

Haochen Tian[*], Defeng Zou, Youjian Song, Minglie Hu

*Ultrafast Laser Laboratory, Key Laboratory of Opto-electronic Information Technology, Ministry of Education, School of Precision Instruments and Opto-electronics Engineering, Tianjin University, Tianjin 300072, China*

*haochentian@tju.edu.cn

**Abstract:** Investigation on the relative phase evolution between two bounded optical solitons is essential for its potential applications in development of larger telecommunication capacity of optical fiber transmission lines, resolution improvement in advancing ultrafast characterization approaches and development of all-optical information storage. Here we characterized relative comb-line phase jitter power spectral density (PSD) of a soliton molecule pair generated from a passively mode-locked Er:fiber laser. Through tracking the intensity difference between two selected wavelengths in one certain spectral interference fringe, the relative comb-line phase noise PSD is measured by balanced detection. The estimated measurement resolution is at $10^{-14}$ rad$^2$/Hz level and the estimated integrated phase noise from 10 MHz to 100 Hz is 2.04 mrad. The estimated relative linewidth is far below 1 mHz. Comparison between phase noise PSD and intensity noise PSD indicates that AM-PM conversion plays an important role in relative phase jitter dynamics between the two solitons. Our spectral interference fringe tracking technique is attractive for its simplicity and shows potential in ultra-high resolution in phase noise measurement, providing an ultra-sensitive alternative approach for relative phase noise stabilization of optical soliton molecules, between two independent mode-locked lasers and optical length stabilization of interferometers.

1. Introduction

Mode-locked lasers have long been applied as practical platforms for the study of complex dissipative nonlinear dynamics widespread in nature. Twin-soliton molecule, which consists of two strongly bounded optical solitons from a mode-locked laser, has attracted special attention due to the central role on the investigation of dynamic attraction behaviors in dissipative systems [1-3]. In particular, study on the relative phase evolution [4-7] between two optical solitons is essential for its potential applications in development of larger telecommunication capacity of optical fiber transmission lines, resolution improvement in advancing ultrafast characterization approaches and all-optical information storage [8].

In-depth investigation into the soliton interaction dynamics based on various dissipative systems requires advanced probing methods. Routinely, observation of relative phase evolution in soliton molecule relies on time-stretch dispersive Fourier-transform (TS-DFT) technique [9]. This technique permits single shot optical spectrum measurement by employing chromatic dispersion-stretched optical pulse to map the broadband spectrum from temporal waveform. Polar diagrams depicting relative phase evolution information have been resolved by this technique in a variety of ultrafast and transient soliton pairs with different soliton separations [10-11]. Recent years, dynamics of distorted and undistorted soliton molecules [12], dissipative optical soliton molecules [13], optical soliton molecular complexes [14], harmonic mode-locking states [15] and other varieties of compound pulsation process [16] have been systematically studied implementing TS-DFT as a real-time probe. Despite TS-DFT technique provides a real-time probing method to observe versatile phase variations in round-trip soliton pair, its capacity of fulfilling high-resolution phase noise measurement is restrained by the finite sampling rate of digital oscilloscope.

In this work, we push the resolution of relative phase noise measurement to a higher level. The relative comb-line phase jitter within a twin-soliton molecule has been characterized with high speed and sub-mrad precision from the intensity difference between two selected wavelengths in an optical spectral interference fringe. The relative comb-line phase noise PSD has been characterized up to 10 MHz Fourier frequency with estimated measurement resolution of $10^{-14}$ rad$^2$/Hz. The estimated integrated phase noise from 10 MHz to 100 Hz is 2.04 mrad and the estimated relative linewidth is far below 1 mHz. Comparison between phase noise PSD and intensity noise PSD indicates that AM-PM conversion plays an important role in relative phase jitter dynamics between the two solitons. Moreover, anti-correlation between relative repetition rate (n×$f_{rep}$) noise PSD, carrier-envelope offset frequency ($f_{ceo}$) noise PSD and comb mode frequency noise PSD is reported for the first time. It should be pointed out that, we admit that the calibration of PSD at high Fourier frequency is not completely rigid, due to the reason that interpolating process is not a strict calibration method and may cause several dB errors on the whole PSD, which would bring uncertainty to the determination of measurement resolution and integrated phase noise. Despite of this, our technique is still attractive for its simplicity and shows superior potential in ultra-high resolution in phase noise measurement. For example, this technique is capable of reducing the residual phase noise in coherent beam combination between two identical mode-locked lasers [17]. Moreover, our noise spectrum analysis provides in-depth insight into ultrafast soliton molecular dynamics, which sets essential fundaments to physical mechanisms in nonlinear optical phenomena, larger optical communications capacity, high-resolution spectroscopy and high-precision phase noise stabilization.

2. Principle

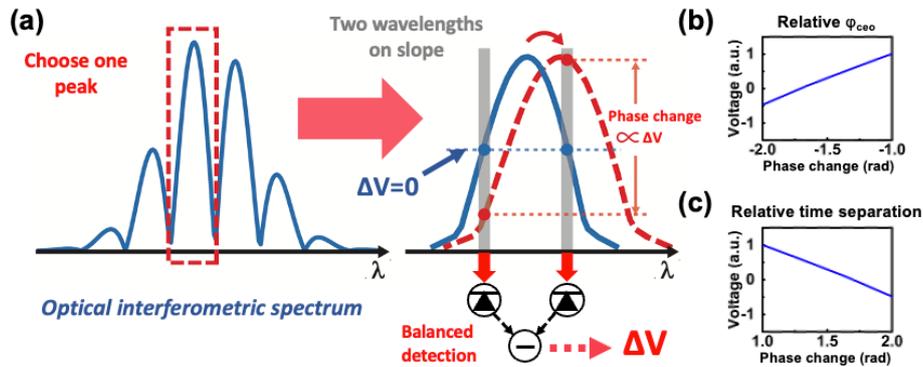

Fig. 1. (a) Principle of phase noise measurement. (b) Voltage variation with change of relative $\varphi_{ceo}$. (c) Voltage variation with change of time separation.

The principle of the phase noise measurement is shown in Fig. 1. The main idea of this approach is borrowed from a carrier-envelope phase noise measurement [18]. Optical spectrum of a soliton molecule pair has obvious interferometric fringes due to the similar optical spectra and low relative phase noise between two optical solitons. Two narrow wavelengths symmetrically on the center of rising and falling edge in one certain interferometric fringe are filtered out, as shown in the grey-shaded part in Fig. 1. Two photo detectors from a balanced photodetector could be applied to detect these two wavelengths' intensity separately. If the intensities of two selected wavelengths are equal, the resulted subtraction of two photodiodes' current is zero. However, when the relative comb-line phase between two solitons drifts, the interferometric fringe changes, leading to an intensity difference between two filtered wavelengths. Therefore, the output voltage from balanced photodetector would deviate from zero voltage. We ran a simple simulation on the output signal from the balanced detector. Through implementing two same pulses with 35-nm FWHM spectral bandwidth and 1.2-ps pulse separation, which is coordinate with the pulse pair under test in our experiment, the output signal from balanced detector varies linearly with time separation change and relative carrier-envelope phase ($\varphi_{ceo}$) change, as shown in Fig. 1 (b) and Fig. 1 (c), respectively. It should be noted that, in these two figures, the phase changes are normalized to 193 THz and voltage changes are normalized by the same factor. This proves that the output of balanced detection has similar response to time separation and relative carrier-envelope phase, leading to the characterization of relative comb mode frequency noise between the two solitons. Moreover, it is beneficial to use balanced detection which is capable of suppressing the intensity fluctuation of the optical spectrum itself, leading to a high-resolution, intensity-drift-free phase noise measurement.

3. **Experimental setup and results**

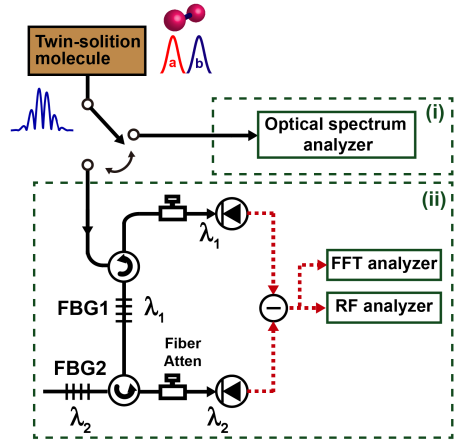

Fig. 2. Experiment setup. FBG, fiber Bragg grating; Fiber Atten: tunable fiber attenuator. (i) scheme of PSD measurement at low Fourier frequency (ii) scheme of PSD measurement at high Fourier frequency

The experiment setup for relative comb-line phase noise PSD measurement of soliton molecules is shown in Fig. 2. The soliton molecule under test is generated from a nonlinear amplified loop mirror (NALM) based Er:fiber mode-locked laser [19]. At single-pulse operation condition, the laser outputs a pulse train with 20 mW average power. The pulse duration is 120 fs. Increase of pump power and slightly tune of wave plate would lead to stable soliton molecule. The output optical spectrum is shown in Fig. 3(a). The low frequency and high frequency phase noise PSD have been characterized separately. For low-frequency noise PSD, we use an optical spectrum analyzer to record the output spectra at a rate of ~ 0.25 Hz over 30 minutes. The recorded evolutive optical spectra is shown in Fig. 3(b). Noise PSDs of relative n×$f_{rep}$, $f_{ceo}$, and comb modes frequency at low frequency (< 100 mHz) could be retrieved from the recorded interference fringes by monitoring the drifts of interferometric peaks' period, the drifts of one certain interferometric peak and intensity changes of a fixed wavelength in the interferometric spectrum, as shown in curve (iii), curve (ii) and curve (i) in Fig. 3 (c), respectively. Here we find that noise PSD of comb mode frequency is not the sum of noise PSD of $f_{ceo}$ and n×$f_{rep}$, where *n* is the mode number. Nevertheless, the fluctuations in $f_{ceo}$ and n×$f_{rep}$ to some extent cancel out each other, resulting a much lower level of comb-line noise. This anti-correlation behavior has been reported in discussed literatures [20-22] in mode-locked lasers. It is appealing to find the similar dynamic in the relative noise of soliton molecules.

For high frequency phase jitter PSD characterization, two reflective FBGs with center wavelengths of $\lambda_1$=1553 nm and $\lambda_2$=1556 nm filter out two narrow wavelengths belonging to the same interference fringe, as shown in Fig. 1. The two wavelengths are tuned to similar average power by fiber attenuators and directed into the detectors of a balanced photodiode detector (Thorlabs PDB420C). The relative comb-line phase changes, either from time separation or relative carrier-envelope phase in the soliton molecule pair result in a proportional differential voltage. Balanced photodetection effectively eliminates common-mode intensity fluctuations as illustrated in the principal section. A fast Fourier transform (FFT) analyzer and a radio frequency (RF) spectrum analyzer are used to measure the phase noise spectrum of the soliton molecule, as shown in Fig. 3(c) curve (iv). This high frequency PSD was calibrated by interpolating curve (i) towards high frequency. The estimated measurement noise floor is at $10^{-14}$ rad$^2$/Hz level. The integrated phase noise from 10 MHz to 100 Hz is 2.04 mrad.

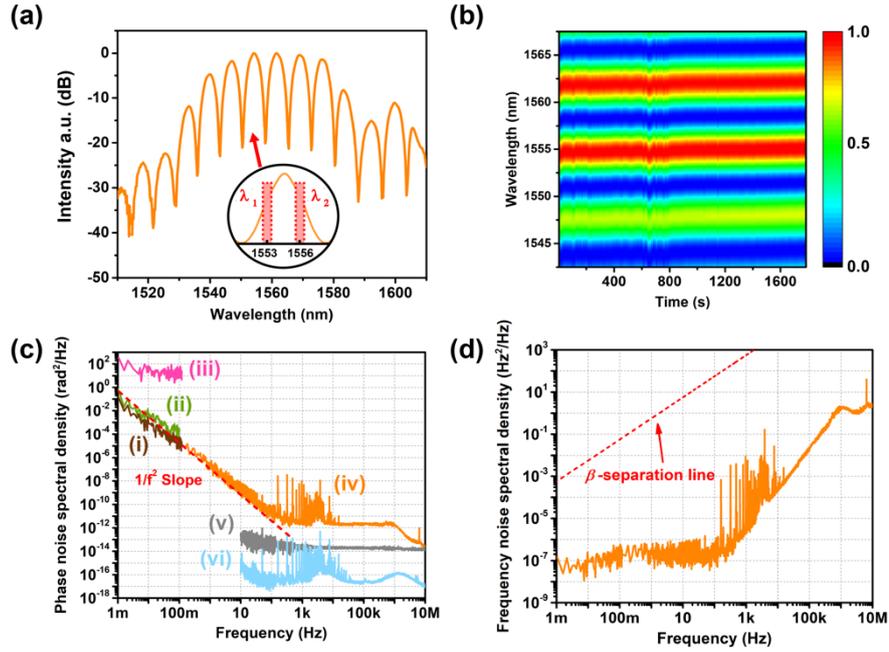

Fig. 3. (a) Spectrum of soliton molecule. Inset shows two selected wavelengths. (b) Recorded output spectra over 30 min. (c) The results of phase noise measurement: Curve (i) represents the relative comb-line noise PSD. Curve (ii) represents the relative $f_{ceo}$ noise PSD. Curve (iii) represents the relative n×$f_{rep}$ noise PSD. Curve (iv) represents the noise PSD measured by balanced detection. Curve (v) represents the photodetection noise floor. Curve (vi) represents RIN induced phase noise during measurement. (d) Frequency noise PSD of the soliton molecule.

Several noise origins need to be considered in order to verify the validity of the phase noise measurement. The first one is the noise generated by the photodetector. For the reason that the input power on each photodiode is less than 1 mW, shot noise should be below the thermal noise of the photodetector. The thermal noise is measured by characterizing the output voltage PSD of balanced photodetector without any input. Through dividing the voltage PSD by the calibration factor of curve (iv), the PSD of thermal noise from the photodetector could be obtained, as shown in curve (v) in Fig. 3(c). The result indicates that our phase noise measurement is thermal-noise-limited. Next, the influence of relative intensity noise (RIN) to the phase noise measurement is estimated. The intensity noise is at ~-130 dBc/Hz level at Fourier frequency from 10 Hz to 1 MHz. Above 1 MHz, the noise PSD starts to roll off. With the function of common mode rejection of the balanced photodetector, which is ~35 dB in our case, the RIN could be reduced to -165 dBc/Hz level. Thus, the impact of RIN to the phase noise measurement is estimated to be at

$(0.5×π)^2×10^{-16.5}=7.80×10^{-17}$ rad$^2$/Hz level, as shown as curve (vi) in Fig. 3 (c). This is ~50 dB below the measured phase noise PSD, making a negligible influence on the phase noise measurement.

The calibration of curve (iv) can't be regarded as 100% strict and rigid. Interpolating process is not a strict calibration method and may cause several dB errors on the whole PSD in high Fourier frequency, bringing uncertainty to the determination of measurement resolution and integrated phase noise. To solve this problem, we tried following two other methods to calibrate. Due to the reason that the relative carrier-envelope phase ($\Delta\varphi_{ceo}$) of soliton pair is quite sensitive to the pumping power and cavity dispersion. The first method is to add pump modulation to the pump diode of the fiber laser. A sinusoidal pump modulation was applied on the pump diode in order to change $\Delta\varphi_{ceo}$ artificially. The interferometric spectrum was supposed to shift 'one fringe' (which refers that peaks turn into gaps while gaps turn into peaks), corresponding to over π change in $\Delta\varphi_{ceo}$. Accordingly, the error signal would reach its minimum ($\Delta\varphi_{ceo} = -1/2$ π) and maximum ($\Delta\varphi_{ceo} = 1/2$ π) on oscilloscope successively. In this way, calibration from error voltage to phase could be fulfilled using the linear range of sinusoidal discrimination signal. Unfortunately, when we were modulating the pump power, the soliton molecule mode-locking state always disappeared before reaching π change of $\Delta\varphi_{ceo}$. Secondly, we made an attempt in cavity dispersion change. It was hard to alter cavity dispersion continuously for an Er:fiber laser without an intra-cavity grating pair or prism pair. To this end, we did this in a Ti:sapphire mode-locked laser with an intra-cavity a prism pair, which emits soliton molecule pair as well [2]. But again, the soliton molecule mode-locking state couldn't be held on before reaching π change of $\Delta\varphi_{ceo}$. As so much statement above, the rigid calibration still remains a technical difficulty to us. We have to admit that, to some extent, our phase noise measurement is only an estimation and more strict calibration method needs to be explored in the future.

However, the general trend of PSD is still convincing. It can be seen from the relative phase noise PSD that, for the Fourier frequency < 1 kHz, the spectrum shows a $1/f^2$ slope characteristic. This slope indicates a random walk nature of the relative phase between the two pulses. For the frequency beyond 1 kHz, the spectrum is dominated by white phase noise and starts to roll off at > 1 MHz, meaning that the phases of the two solitons are only bounded at > 1 MHz bandwidth. The spikes at ~ kHz arise from acoustic vibrations. The RIN spectrum has the similar acoustic spikes at ~ kHz and rolling off features at > 1 MHz, as shown as curve (vi) in Fig. 3 (c). This indicates that AM-PM conversion plays an important role in relative comb-line phase jitter dynamics between the two solitons. Linewidth estimation is realized after converting the phase noise spectrum to frequency noise spectrum, as shown in Fig. 3(d). It can be seen from the $\beta$-separation line that, the relative linewidth of soliton molecule pair is far below 1 mHz [23]. Assuming $1/f^2$ slope along lower frequency, the relative linewidth is estimated to be at 1 μHz level.

## 4. Conclusion and discussion

In conclusion, we characterized the relative comb-line phase PSD of a temporal soliton molecular for the first time with high resolution. The relative phase noise PSD has been characterized up to 10 MHz Fourier frequency with estimated measurement resolution of $10^{-14}$ rad$^2$/Hz. The estimated integrated phase noise from 10 MHz to 100 Hz is 2.04 mrad and the estimated relative linewidth is far below 1 mHz. Comparison between phase noise PSD and intensity noise power spectral density reveals that AM-PM conversion plays an important role in relative phase jitter dynamics between the two solitons. Despite that the calibration of PSD at high Fourier frequency is not rigid, our spectral interference fringe tracking technique still shows ultra-high resolution in phase noise measurement. Compared with well-established TS-DFT technique, our phase noise measurement method gets rid of long-term drift of refractive index from kilometer-long fiber [24]. In the meanwhile, our method pushes measurement resolution to about ~40 dB higher than TS-DFT technique, where the resolution is mainly limited by finite sampling rate [25]. This study provides a new perspective towards the soliton molecular dynamics and may contribute to the development of more rigorous models of complex nonlinear systems. Moreover, spectrum analysis of noise dynamics is motivated for both shedding new light on the detailed nature of fundamental physics towards the ultrafast nonlinear optics and extending potential applications of larger telecommunication in optical fiber transmission lines, real-time spectroscopy, etc.

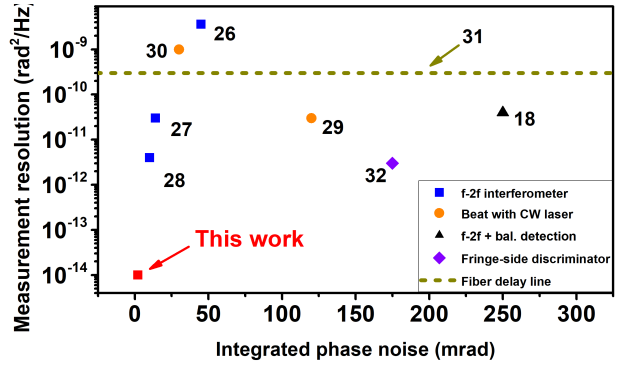

Fig. 4. Summary of resolution in representative phase noise measurements.

On the other side, compared with other carrier-envelope phase and comb-line phase noise measurements in mode-locked lasers and CW lasers, the estimated sensitivity of our method shows significant potential to exceed the sensitivity of phase noise measurements in discussed literatures [18, 26-32], as compared in Fig. 4. In practical, our high-sensitivity phase noise detection technique could be further applied in relative comb-line phase noise stabilization of the two bounded solitons. We estimated that, if a 100-kHz bandwidth pump power modulation could be implemented, the resulting residual phase noise could be pushed down to < 1.9 mrad. Not restrained to the research of soliton molecule, this presented technique can be also implemented in coherent beam combinations [17] and optical length stabilization of interferometers due to its high detection sensitivity.

**Acknowledgments**

This work was supported in part by National Natural Science Foundation of China (NSFC) (61975144, 61827821, 61675150). The authors thanks Profs. Thomas Schibli in University of Colorado, Profs. Kaoru Minoshima in University of Electro-Communications and Yu Cai in Tianjin University for their helpful discussion.

**References**

[1] P. Grelu, and N. Akhmediev. "Dissipative solitons for mode-locked lasers." Nature photonics, 6, 2 (2012): 84.

[2] Y. Song, et al. "Attosecond timing jitter within a temporal soliton molecule." Optica, 7, 11 (2020): 1531.

[3] H. Shi, et al. "Observation of subfemtosecond fluctuations of the pulse separation in a soliton molecule." Optics Letters, 43, 7 (2018): 1623.

[4] N. N. Akhmediev, A. Ankiewicz, and J. M. Soto-Crespo. "Multisoliton solutions of the complex Ginzburg-Landau equation." Physical Review Letters, 79, 21 (1997): 4047.

[5] V. V. Afanasjev, B. A. Malomed, and P. L. Chu. "Stability of bound states of pulses in the Ginzburg-Landau equations." Physical Review E, 56, 5 (1997): 6020.

[6] B. A. Malomed. "Bound solitons in the nonlinear Schrödinger–Ginzburg-Landau equation." Physical Review A, 44, 10 (1991): 6954.

[7] A. Zavyalov, et al. "Discrete family of dissipative soliton pairs in mode-locked fiber lasers." Physical Review A, 79, 5 (2009): 053841.

[8] M. Pang, et al. "All-optical bit storage in a fibre laser by optomechanically bound states of solitons." Nature Photonics, 10, 7 (2016): 454.

[9] K. Goda, and B. Jalali. "Dispersive Fourier transformation for fast continuous single-shot measurements." Nature Photonics, 7, 2 (2013): 102.

[10] G. Herink, et al. "Real-time spectral interferometry probes the internal dynamics of femtosecond soliton molecules." Science, 356, 6333 (2017): 50.

[11] K. Krupa, et al. "Real-Time Observation of Internal Motion within Ultrafast Dissipative Optical Soliton Molecules." Physical Review Letters, 118 (2017): 243901.

[12] J. Igbonacho, et al. "Dynamics of distorted and undistorted soliton molecules in a mode-locked fiber laser." Physical Review A, 99, 6 (2019): 063824.

[13] J. Peng and H. Zeng. "Build-Up of Dissipative Optical Soliton Molecules via Diverse Soliton Interactions." Laser & Photonics Reviews, 12, 8 (2018): 1800009.

[14] Z. Q. Wang, et al. "Optical soliton molecular complexes in a passively mode-locked fibre laser." Nature communications, 10, 1 (2019): 1.

[15] X. Liu and M. Pang. "Revealing the Buildup Dynamics of Harmonic Mode-Locking States in Ultrafast Lasers." Laser & Photonics Reviews, 13, 9 (2019): 1800333.

[16] Z. Wang, et al. "Self-organized compound pattern and pulsation of dissipative solitons in a passively mode-locked fiber laser." Optics Letters, 43, 3 (2018): 478.

[17] H. Tian, et al. "Long-term stable coherent beam combination of independent femtosecond Yb-fiber lasers". Optics Letters, 41, 22 (2016): 5142.

[18] A. Liehl, et al. "Ultrabroadband out-of-loop characterization of the carrier-envelope phase noise of an offset-free Er: fiber frequency comb." Optics Letters, 42, 10 (2017): 2050.

[19] R. Li, et al. "All-polarization-maintaining dual-wavelength mode-locked fiber laser based on Sagnac loop filter." Optics Express, 26, 22 (2018): 28302.

[20] V. Dolgovskiy, et al. "Cross-influence between the two servo loops of a fully stabilized Er: fiber optical frequency comb." JOSA B, 29, 10 (2012): 2944.

[21] P. Brochard, et al. "Frequency noise characterization of a 25-GHz diode-pumped mode-locked laser with indirect carrier-envelope offset noise assessment." IEEE Photonics Journal, 10, 1 (2017): 1-10.

[22] H. Tian, et al. "Optical frequency comb noise spectra analysis using an asymmetric fiber delay line interferometer." Optics Express, 28, 7 (2020): 9232.

[23] G. Di Domenico, S. Schilt, and P. Thomann. "Simple approach to the relation between laser frequency noise and laser line shape." Applied Optics, 49, 25 (2010): 4801.

[24] K. Goda and B. Jalali. "Dispersive Fourier transformation for fast continuous single-shot measurements." Nature Photonics, 7, 2 (2013): 102.

[25] A. Mahjoubfar, et al. "Time stretch and its applications." Nature Photonics, 11, 6 (2017): 341.

[26] S. Koke, et al. "Direct frequency comb synthesis with arbitrary offset and shot-noise-limited phase noise." Nature Photonics, 4, 7 (2010): 462.

[27] T. D. Shoji, et al. "Ultra-low-noise monolithic mode-locked solid-state laser." Optica, 3, 9 (2016): 995-998.

[28] R. Liao, et al. "Active f-to-2f interferometer for record-low jitter carrier-envelope phase locking." Optics Letters, 44, 4 (2019): 1060.

[29] L. C. Sinclair, et al. "Invited Article: A compact optically coherent fiber frequency comb." Review of Scientific Instruments, 86, 8 (2015): 081301.

[30] N. Kuse, et al. "All polarization-maintaining Er fiber-based optical frequency combs with nonlinear amplifying loop mirror." Optics Express, 24, 3 (2016): 3095.

[31] C. Li, et al. "All-optical frequency and intensity noise suppression of single-frequency fiber laser." Optics Letters, 40, 9 (2015): 1964.

[32] N. Coluccelli, et al. "Frequency-noise measurements of optical frequency combs by multiple fringe-side discriminator." Scientific Reports, 5 (2015): 16338.